\documentclass{svjour3}
\smartqed  

\usepackage{url}
\usepackage{amssymb}
\usepackage{amsmath}
\usepackage{multirow,multicol}
\usepackage{tabularx}
\usepackage[table]{xcolor}
\usepackage{subfig,dblfloatfix}
\usepackage{graphicx}
\usepackage{rotating}

\usepackage{tcolorbox}
\tcbset{colback=gray!10,colframe=gray!60,boxrule=0.6pt,arc=2pt,left=6pt,right=6pt,top=4pt,bottom=4pt}

\newcommand{\PreprintNote}{%
\vspace{-0.5em}
\begin{tcolorbox}
\footnotesize
\textbf{Preprint (not peer-reviewed).}
The peer-reviewed Version of Record is available at:
\texttt{https://doi.org/10.1007/s11042-018-6641-x}.
\end{tcolorbox}
\vspace{-0.5em}
}

\begin{document}

\title{Visualizing a Collective Student Model for Procedural Training Environments}

\author{D. Riofr\'{i}o-Luzcando \and
        J. Ram\'{i}rez \and
        C. Moral \and
        A. de Antonio \and
        M. Berrocal-Lobo
}

\institute{D. Riofr\'{i}o-Luzcando \at
              International University SEK, Ecuador \\
              Tel.: +593-987640852\\
              \email{diego.riofrio@uisek.edu.ec}
           \and
		J. Ram\'{i}rez \at
        ETS de Ingenieros Inform\'{a}ticos, UPM, Spain
        \and
        C. Moral \at
        ETS de Ingenieros Inform\'{a}ticos, UPM, Spain
        \and
        A. de Antonio \at
        ETS de Ingenieros Inform\'{a}ticos, UPM, Spain
        \and
        M. Berrocal-Lobo \at
        E.T.S.I. Forestal y del Medio Natural, UPM, Spain
        }
\date{Received: January 2018 / Accepted: date}
\maketitle
\PreprintNote

\begin{abstract}
Visualization plays a relevant role for discovering patterns in big sets of data. In fact, the most common way to help a human with a pattern interpretation is through a graphic. In 2D/3D virtual environments for procedural training the student interaction is more varied and complex than in traditional e-learning environments. Therefore, the visualization and interpretation of students' behaviors becomes a challenge. This motivated us to design the visualization of a collective student model built from student logs taken from 2D/3D virtual environments for procedural training. This paper presents the design decisions that enable a suitable visualization of this model to instructors as well as a web tool that implements this visualization and is intended: to help instructors to improve their own teaching; and to enhance the tutoring strategy of an Intelligent Tutoring System. Then, this paper illustrates, with three detailed examples, how this tool can be used to those educational purposes. Next, the paper presents an experiment for validating the utility of the tool. In this experiment we show how the tool can help to modify the tutoring strategy of a 3D virtual laboratory. In this way, it is shown that the proposed visualization of the model can serve to improve the performance of students in 2D/3D virtual environments for procedural training.
\end{abstract}


\section{Introduction}\label{sec:introduction}
Analyzing data has been a hot research topic in the last two decades, mainly since the analysis of huge amounts of data (the so called data mining, or big data) has become essential to face new challenges and solve major problems. Visualization plays a relevant role for this purpose as it can be of great help to visually discover information that may be derived from the data, to guide or validate results obtained from statistical analysis of data, or to recognize clusters and/or patterns \cite{Shneiderman1996}, what is called as visual analytics and is used to discover interesting behaviors from educational data \cite{Vatrapu2011,Aguilar2014}

In a previous work \cite{Riofrio-Luzcando2017}, we presented a collective student model that was designed to anticipate the actions that students are likely to take while completing a practical assignment in a 2D/3D virtual environment for procedural training. This model is created from activity records or logs collected from students with a similar background that previously completed the same practical assignment, and it takes the form of extended automata associated to student clusters. 

The most common way to help humans with pattern interpretation is through graphics. This data mining technique is known as distillation of data for human judgment \cite{Baker2010}, and its objective is to present relevant information in a summarized, interactive and visual form, in order to support decision-making.

Our goal was the design of a visualization to help instructors in the identification of relevant patterns in students' behavior (common errors, difficult tasks, tasks that cause process deadlocks, etc.) from the automaton generated for each cluster of the collective student model. Automata are drawn as graphs, because graphs are useful to represent sequential patterns \cite{Tan2013}. As a result of this, one of the main obstacles that we tackled was how to visualize a possibly huge graph to the eyes to an instructor without cognitively overloading him/her.

By taking advantage of this visualization the instructor can improve his/her teaching to bridge the knowledge gaps revealed by the behavior patterns. Additionally, the discovery of these patterns could help to enhance the tutoring strategy of an Intelligent Tutoring System for a procedural training environment.

In addition, note that in 2D/3D virtual environments for procedural training the student interaction is more varied and complex than in traditional e-learning environments. This fact introduces at the same time a novel aspect and a difficulty in our work, as we will demonstrate later on.

The structure of the remainder of the paper is as follows. Section \ref{sec:related_works} shows relevant works in the visualization of educational data. Section \ref{sec:model} outlines the collective student model to be visualized. Additional details about this model can be found in  \cite{Riofrio-Luzcando2017}. Section \ref{sec:design} describes the design of the proposed visualization and explains the main design decisions related to the representation of nodes and edges and the interaction with the graph. Section \ref{sec:examples} details three application examples of the viewer that implements the proposed visualization. Section \ref{sec:validation} reports an evaluation of the viewer utility. Section \ref{sec:conclusion} outlines the conclusions of this research. Finally, Section \ref{sec:futurework} poses some future works that may be derived from this work.

\section{Related Work}\label{sec:related_works}
Much research has been conducted to identify what are the main aspects that have to be considered to properly visualize information.  Bertin \cite{Bertin1981} was the first author defining the pillars of information visualization, including the marks, that is, the graphical objects used to represent the data in the visualization, and their positional, temporal and retinal properties, such as color (in terms of value, hue, and saturation), transparency, width, curvature, flicker, and size, among others \cite{MacEachren1995,Healey1996}. Some years later, Mackinlay \cite{Mackinlay1986} completed this definition by adding the enclosure and connection properties as a means of illustrating relationships between the marks. Card and Mackinlay \cite{Card1997} have summarized all the visual features into four main categories: spatial features, retinal properties, relationships between data, and animation. More recently, \cite{Moral2016} has proposed a complete UML-based model representing all these aspects.

However, even if all these concepts are valid regardless of the type of data being visualized, there is some consensus in stating that information visualization is domain-dependent \cite{Einsfeld2006}. In the case of educational data, there are some works that have proposed specific information visualization systems to illustrate user models, communications occurred in on-line courses and student's tracking data \cite{Mazza2010}. 

Some tools allow to visualize patterns of failure or success in an on-line course \cite{Jordao2014,Wortman2007}, while others (like AVOJ \cite{Xiaohuan2013}) group students by their grades and study habits to allow teacher to compare these groups. 

Graphical representations of student models are used to display the knowledge each student has acquired during a learning activity and can be used by instructors to adapt their teaching to the beliefs, needs and preferences of the students \cite{Mazza2004}. Some of these representations take the form of hierarchies or trees (e.g. \textit{ConceptLab} \cite{Zapata-Rivera2000}), graphs or networks (e.g. \textit{VisMod} \cite{Zapata-Rivera2001}, and \textit{VIUM} \cite{Uther2003}), progress bars (e.g. \textit{E-Kermit} \cite{Hartley2002}, and \textit{NORMIT} \cite{Mitrovic2002}), map-based visualizations (e.g. \textit{Maths Island Tutor} \cite{Grawemeyer2015}), simple-skill meters (e.g. \textit{LATUX} \cite{Martinez-Maldonado2015}), animations of a learner’s programming code execution (e.g. \textit{AniMis} \cite{Johan2009}), bundle visualizations, tag clouds, or other interactive diagrams (e.g. \textit{TUT LA tool} \cite{Kuosa2016}).

More recently, some visual systems help teachers to understand the interactions of students in the learning environment. For this, authors present the average of interactions by learning level, impact of resources and the influence of interactions for learning \cite{Paiva2018}. Other form is representing the interaction by date and activity \cite{Wan2017}, where in x-axis is placed the date and in y-axis each activity, each interaction on an activity is represented by a spot, the size of the spot represents how many times have the student done the activity, the interaction for homeworks or test submissions are represented by trapezoids, with the start and the end of the assignment and the grade (size of the trapezoid).

Some developed tools offer different visualization that can be selected according to the user’s (learner’s or teacher’s) preferred way of accessing the learner model data such as skill meters, table, treemap, competency network and word clouds \cite{Bull2016}; dashboard with learning records, engagement level in material, watching pattern for a video and assessment results \cite{Lin2017}; dashboard to visualize student engagement in terms of social interaction and use of knowledge objects \cite{Pesare2016}. In addition, other works have explored  the value of adding social dimension to open student models by allowing  students  to  explore  visually models  of  peer  students  or  the  whole  class \cite{Brusilovsky2015}.

Other works aim to graphically visualize communication exchanges in discussion forums, e-mail, or chats, between students and with teachers. Some visualize this kind of information as graphs \cite{Reffay2003}, where nodes represent individuals or groups of individuals and transitions display communications between them. On the other hand, \textit{PeopleGarden} \cite{Xiong1999} uses the flower and garden metaphors to illustrate student participation in a message board, where each flower symbolizes an individual. Moreover, \textit{CourseVis} \cite{Mazza2007} is a system that transforms student social activity data into graphic representations like spheres, bars or points. For example, student discussions in a forum are represented through spheres that are spatially located in a three-dimensional environment based on their author, the date they were published, and the topic they deal with (each of these aspects is mapped onto one of the dimensions of the 3D space). Besides, the size of each sphere is directly related to the number of follow-ups in the represented discussion. Later on, based on \textit{CourseVis}, a Moodle plug-in named \textit{GISMO} \cite{Mazza2007a} was implemented.

In relation to the visualization of big data in education, Li et. al. \cite{Li2015} propose a visual analytics system for e-learning, which can help educators to improve their teaching by presenting them with a vision of the learning process. This system includes two elements: one is a group analysis component, which depicts factors of the student process in a collective form, for example, score vs number of submissions per student; the other is a case analysis component, which can show data detailed for each student. This data can be used to adjust educational strategies to outliers. The case analysis component implements a visualization technique called Bubble Sets map, which shows the number of submission (points) for each exercise (bubbles). A new version of this work \cite{Li2017} is used to analyze MOOCs data.

Although there has been substantial interest in learning analytics regarding the visualization of educational data \cite{Klerkx2014,Tervakari2014}, to the best of our knowledge, there are not previous works that have addressed the visualization of data collected by tracking the actions performed by the students in a 3D/2D virtual learning environment for procedural training. Moreover, the existing visualization proposals explore different aspects of the interaction of the student with the learning environments (such as communications flows, number of submissions, watching patterns or assessment results, etc.) that are not applicable for the visualization of student actions in 3D/2D virtual learning environments for procedural training.

Thus, the existing gap in the state of the art motivated the work presented in this paper so that it describes a first step forward to bridge this gap.

\section{The Collective Student Model} \label{sec:model}

The collective student model \cite{Riofrio-Luzcando2017} to be visualized comprises summarized data on past student action events (logs). This will be used to predict the actions that a student under supervision is most likely to take next. The premise for creating this model is that the behavior of current students will be similar to that of past students with a similar background completing the same practical assignment. 

This model is built using logs from past students created by an intelligent tutor that supervises the behavior of the students while they are performing the practical assignment in a virtual environment for procedural training.

The model groups students into several clusters, each of which contains an extended automaton. The purpose of dividing students into groups is to provide an automatic tutoring that can adapt itself to different student types.

As is explained in a previous work \cite{Riofrio-Luzcando2017}, this model was validated using some clustering methods, XMeans, Expectation Maximization (EM), Microsoft Sequence Clustering, No clustering. First two methods were used with the following clustering functions:
\begin{itemize}
\item Clustering by errors
\item Clustering by errors and time
\item Clustering by events in each zone of the automaton.
\end{itemize}

The first function returns an error coefficient, which is calculated using the weighted sum of errors made by a student with a different weight depending on the pedagogical importance of the error. The second function returns an error coefficient-time pair, where the error coefficient is calculated as explained before, and time stands for the time that it took the student to complete the entire practical assignment. The clustering by events in zone function returns a triple with the number of events of each type existing in a student log.

In this validation we concluded that for our model and with the data we used, XMeans with the events by zone clustering function can successfully classify students into groups that will need different tutoring feedback throughout the practical assignment \cite{Riofrio-Luzcando2017}.

\subsection{Extended Automaton Definition}\label{sec:automaton}
\begin{figure*}[h]
 \centering
 \includegraphics[width=\textwidth]{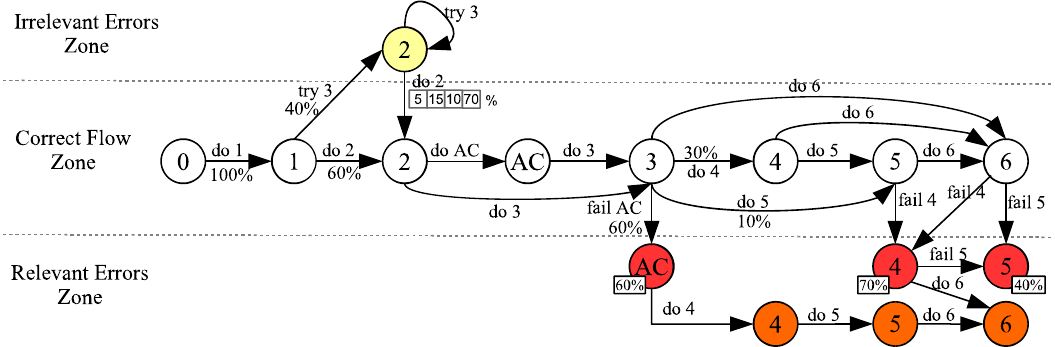}
 \caption{Example of the automaton of a simple activity}
 \label{fig:example_automaton}
\end{figure*}
An example of an automaton of a cluster is displayed in Figure \ref{fig:example_automaton}, where states are represented by circles and transitions by arrows. Transitions denote events generated by students throughout an exercise. An event in the automaton can represent one of the following situations:

\begin{itemize}
\item a valid action for an exercise (do X events);
\item an attempted action blocked by the intelligent tutor because it is wrong and the tutoring strategy has been configured to prevent students from performing this action (try X events); or
\item an error detected by the intelligent tutor at the time of validating an incorrect action that has not been blocked (fail X events).
\end{itemize}

Therefore, states represent the different situations derived from the events generated by students.

States are grouped into three possible categories: 
\begin{itemize}
\item Correct Flow: states that are reached with valid sequences of actions for an exercise.
\item Irrelevant Errors: states derived from error events that do not influence the final result.
\item Relevant Errors: states derived from error events that influence the final result, including states derived from correct events that come after an error.
\end{itemize}

Each state contains the number of students whose logged sequences of events have passed through that state. Likewise, each transition also contains its student frequency, which is the number of logged sequences of events that have passed through the transition.

Figure \ref{fig:example_automaton} shows how some events associated with right actions recorded in student logs produce correct states. For example, event ``do 1'' leads to state 1 with frequency 100\%, which means that all the students performed correctly that action. In contrast, the event that leads to yellow state 2 does not represent a right action, but the error event of trying to do action 3 instead of action 2.

For pedagogical reasons, the virtual environment typically makes more actions available to students than those required for completing the practical assignment. Some of these actions may be wrong and to represent that, they would be configured as incompatible with other right actions. That is the case of action AC, which is considered incompatible with the right action 3 in Figure \ref{fig:example_automaton}. Hence, as soon as action 3 is performed, the ``do AC'' event is validated as incorrect producing a ``fail AC'' event that leads to a red AC state. As is shown in Figure \ref{fig:example_automaton}, some students completed the remainder of the practical assignment correctly after performing ``do AC'' (path of orange states leading to orange state 6).

As regards relevant errors, red state 4 is caused by the fail event of not doing the right action 4 after action 3, because some students performed action 5 too early. For example, the automaton represents the fact that some students mistakenly performed action 5 instead of action 4 by the path: \(do~3 \to do~5 \to fail~4\). In addition, a wrong action can cause more than one error event at the same time in a student log. For example, if some students mistakenly perform action 6 from state 3 instead of action 4, it will be represented by the path: \(do~6 \to fail~4 \to fail~5\), because students have skipped two consecutive right actions.

\section{Design of the Visualization} \label{sec:design}

In the design of this visualization, we had to address several problems that arise when visualizing a graph of large dimension \cite{Mazza2009}:

\begin{itemize}
\item Position and representation of the nodes.
\item Representation of the edges.
\item Dimensioning the graph.
\item Interaction with the graph.
\end{itemize}

\begin{figure*}[!h]
 \centering
\includegraphics[width=\textwidth,height=8cm]{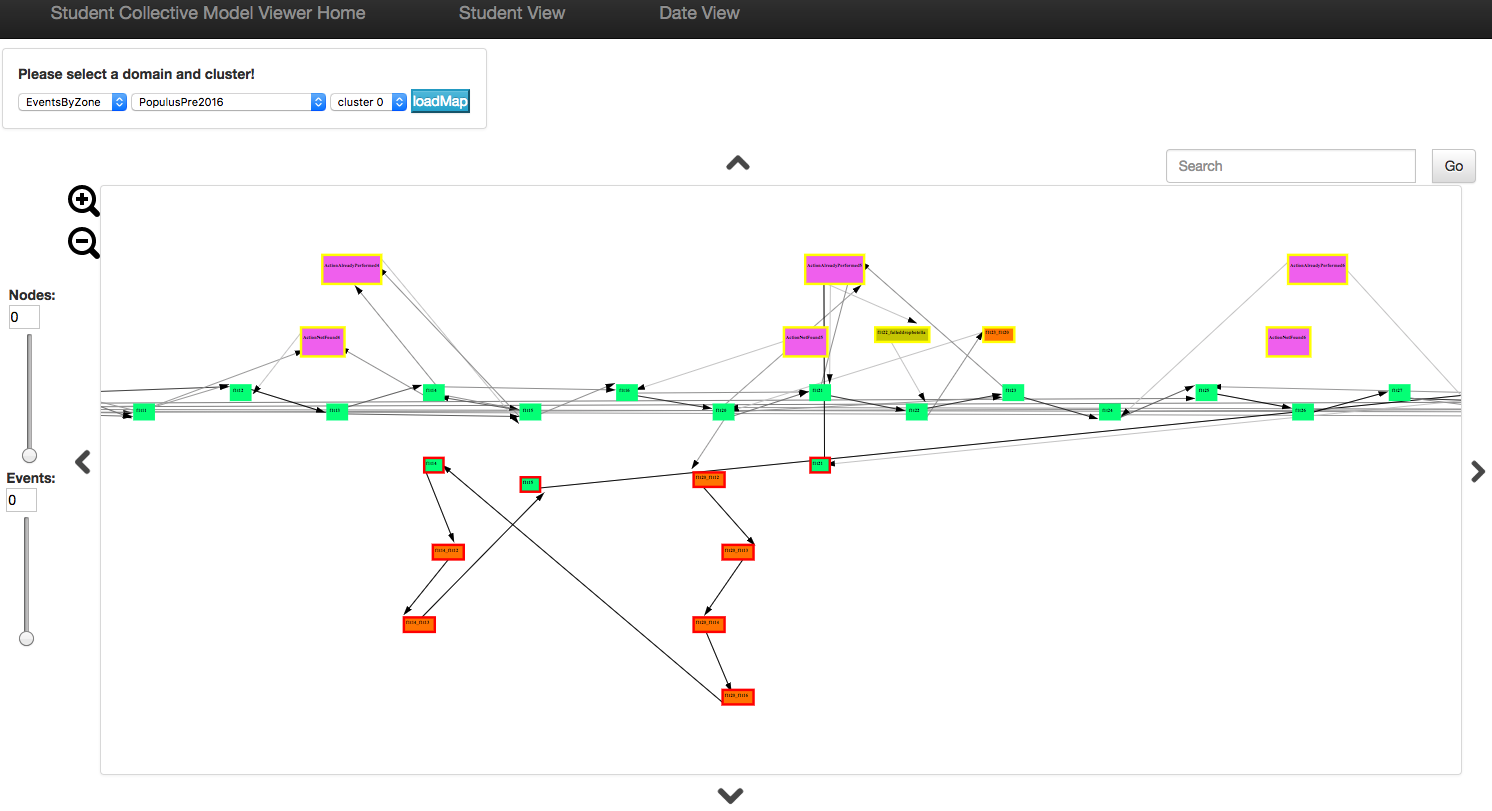}
 \caption{Screenshot of the collective student model viewer.}
 \label{fig:visor}
\end{figure*}

In the next sections, we will explain how each of these problems were solved. 

Before beginning with the design of the visualization, it is worth clarifying that all the graphs could not be visualized in the same way as the graph of Figure \ref{fig:example_automaton}. This is because of different reasons, but some of them would be that in a large graph: there are many nodes and edges, therefore if we drew also information on frequencies or event names, we would easily overload users with too many details and harm clarity of the image; and there are many long edges, therefore it would be pointless to draw a label with the name of the event close to each edge. 

\subsection{Position and representation of the nodes}
Graphs usually represent abstract data, without a natural spatial location. For this reason, it is important to choose a good criterion to distribute the nodes on the drawing space, depending on the application domain \cite{Mazza2009}. One possible solution to this problem is to use ``spatial'' variables associated with nodes \cite{Card1999}. 

Taking into account that the states in our model \cite{Riofrio-Luzcando2017} are located abstractly in different categories, this property has been considered as the ``spatial'' characteristic of each node. Hence, we decided to divide the drawing area into three zones corresponding to the categories detailed in section \ref{sec:automaton}, and states are positioned on the zone to which they belong, as Figure \ref{fig:visor} shows. 

The correct flow zone is positioned in the center, because all students begin their training from a state located in this zone. Later the sequence of events can be directed to other zones depending on the errors committed by the student. In addition, as states are reached sequentially in the definition of the model \cite{Riofrio-Luzcando2017}, the most obvious solution to the problem of positioning states is to draw them consecutively according to the chronological order of the events happening in the training process. 

However, this solution works only for states in the correct flow zone, because these states will be ordered always in a strict sequence. States in the irrelevant errors zone are located above the correct flow zone, similarly to states in relevant errors zone which are located below the correct flow zone. These error states are aligned vertically with respect to the correct state where the error events come from. This layout facilitates understanding the situations where students make mistakes.

It can be distinguished in the Figure \ref{fig:visor} that states are not perfectly aligned horizontally or vertically. This distribution was implemented to avoid the overlapping of edges that connect states that are in different positions of the sequence.

\begin{figure}[htb]
 \centering
\includegraphics[width=2cm,height=1.5cm]{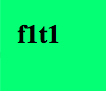}
 \caption{Example of a state reached after performing the action f1t1.} 
\label{fig:visor_forma_estado}
\end{figure}

At the moment of graphing the states there are other relevant characteristics to be considered, such as their shape, color and size. Since this model resembles a conceptual map \cite{Mazza2009}, it is necessary to include within each state the code of the event that generated it. Therefore, a rectangular shape has been adopted for the nodes (Figure \ref{fig:visor_forma_estado}), because it is the best shape that fits a text box. The width of the nodes is automatically set depending on the size of the text to be included. The codes of the events come from the configuration of the practical assignment in the Intelligent Tutor. therefore the professor should know them. However, if the user needs more meaningful information on a node, he/she only has to click on the state to obtain more details that allow the user to find out what the state stands for.

States are also distinguished by their filling and contour color. The filling color differentiates states by their type, and the contour color by the zone they belong to. In this way, apart from the vertical location of the state, the zone to which a state corresponds is also distinguished by its contour color, according to the following color coding:
\begin{itemize}
\item Correct Flow Zone \begin{tabular}{c}\cellcolor[RGB]{4,255,117}\end{tabular} (green).
\item Irrelevant Errors Zone \begin{tabular}{c}\cellcolor[RGB]{255,255,3}\end{tabular} (yellow).
\item Relevant Errors Zone \begin{tabular}{c}\cellcolor[RGB]{255,0,1}\end{tabular} (red).
\end{itemize}

Depending on the type of state, a different filling color will be used. The employed range of colors is as follows:
\begin{itemize}
\item Correct states \begin{tabular}{c}\cellcolor[RGB]{4,255,117}\end{tabular} (green).
\item Simple dependence error states \begin{tabular}{c}\cellcolor[RGB]{255,128,0}\end{tabular} (dark orange).
\item Complex dependence error states \begin{tabular}{c}\cellcolor[RGB]{255,153,0}\end{tabular} (orange).
\item Incompatibility error states \begin{tabular}{c}\cellcolor[RGB]{215,104,89}\end{tabular} (red orange). 
\item Time error states \begin{tabular}{c}\cellcolor[RGB]{255,255,0}\end{tabular} (yellow).
\item World error states \begin{tabular}{c}\cellcolor[RGB]{204,204,0}\end{tabular} (dark yellow).
\item Action already performed and action not found error states \begin{tabular}{c}\cellcolor[RGB]{241,106,239}\end{tabular} (light purple).
\end{itemize}

Dependence, incompatibility and time errors depend on the configuration of the virtual environment set up by the instructor. Instead, world errors refer to failures in the handling of 2D/3D objects, for example, if a student tries to drop an object where it should not be dropped. Therefore, world errors are defined as part of the design of the interaction with the virtual environment. Dependence errors are related to the right order in which to perform the actions in the practical assignment. Incompatibility errors are the result of the performance of actions that should not have been done before other actions. Time errors occur when time constraints for performing some action are not respected, i.e. when the student performs some action too late or too early after doing other action.

\begin{figure}[htb]
 \centering
\includegraphics[width=4cm]{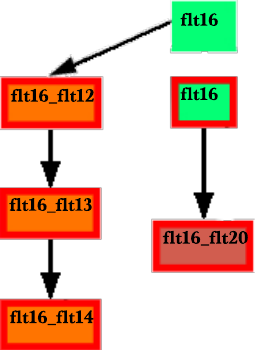}
 \caption{Layout of relevant error states from a correct state.} 
\label{fig:visor_erroresvertical}
\end{figure}

As can be observed in Figure \ref{fig:visor_erroresvertical}, some simple dependence error states (orange states with red outline) are hanging from state f1t16. This is because the action that produces this event (action f1t16) has some preconditions or dependencies, i.e., actions that must be carried out before action f1t16; and these preconditions were not met by some students, because they had not previously executed the actions f1t12, f1t13 and f1t14.

Another example of relevant error state in Figure \ref{fig:visor_erroresvertical} is the state of an incompatibility error (red orange with red outline). It is  hanging from a correct state in the relevant error zone (green with red outline). This correct state is derived from the same action f1t16 of the previous example, but as it drags a previously committed error, it is in the area of relevant errors.

If the automatic tutor finds errors of the two types when it is validating an action, we will draw all the state errors hanging from the state reached when performing this action, in the following top-down order: first dependence error states and then incompatibility error states.

\textit{Action already realized} states represent error states reached after performing the same action more than once. \textit{Action not found errors} states also stands for error states reached after performing an action that is not expected to be executed in the current phase of the practice, but in a different phase.

\begin{figure}[h]
\centering
\subfloat[]{\label{fig:visor_accionyarealizada}\includegraphics[width=3.5cm,height=1.5cm]{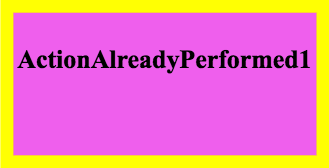}}\quad
\subfloat[]{\label{fig:visor_accionnoencontrada}\includegraphics[width=2.5cm,height=1.5cm]{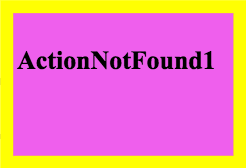}}
\caption{Super-states examples.}
\end{figure}

In a preliminary version of the visualization we used a different \textit{Action already realized} state for each action. However, we realized that we could facilitate the understanding of the graph in the irrelevant errors zone by grouping each subsequence of N consecutive states of the type \textit{action already realized} (Figure \ref{fig:visor_accionyarealizada}) into a single \textit{super-state}. Likewise, consecutive \textit{action not found errors} states were grouped into another \textit{super-state} (Figure \ref{fig:visor_accionnoencontrada}). We were able to adopt this simplification without fear of losing  information relevant to the teacher, because, in general, according to the instructor, these types of errors  have no pedagogical importance. On the other hand, if this type of group states has a high frequency, this could indicate that students are disoriented or having difficulties with the interaction with certain objects in the virtual environment.

\subsection{Representation of the edges}
\begin{figure}[h]
 \centering
\includegraphics[width=0.4\textwidth,height=1.5cm]{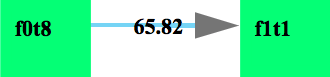}
 \caption{Visualization of an event (edge).\\Frecuency=65.82} 
 \label{fig:visor_evento}
\end{figure}

The frequencies of the events are represented as weights in the edges (see Figure \ref{fig:visor_evento}). Also, the edges are painted following a scale of grays so that the greater the frequency, the darker will be the arrow.

To lighten the amount of information that is presented on the drawing area at the same time, the edges weights are not presented by default, but each edge weight is shown if the user hovers over this edge. 

\subsection{Dimensioning the graph}
Some data sets have a large number of records, which complicates the representation of all these data simultaneously on the screen by means of a graph \cite{Mazza2009}. We decided that it is preferable to visualize only a part of the graph instead of the graph as a whole, whenever it is too big. 

In this way, the initial view of the automaton \cite{Shneiderman1996} will depend on the size of the browser window. However, the user is allowed to perform a vertical and horizontal panning through the automaton using the buttons on the sides of the drawing area.

\subsection{Interaction with the graph}
To design the interaction with the automaton we follow the taxonomy TTT of Shneiderman \cite{Shneiderman1996}, which defines several tasks that users should be able to perform in an information visualization environment regardless of the type of data presented. Among these tasks, we decided to cover the following ones:

\subsubsection{Zoom}
Zoom can be achieved in two ways, one using the magnifying glass buttons on the left side of the drawing area, and another using the mouse wheel.

According to Shneiderman \cite{Shneiderman1996} users typically have interest in one or more objects in the collection, therefore it is necessary to implement a tool that allows the user to focus on these objects of interest. Taking this into account, our design considers another form of focus consisting of a search box located at the top right. When the code of an action in the training process is entered, followed by the name of the zone, a search is performed for the state associated to this action and the current view of the graph is changed  to the found state.

\subsubsection{Filter}
Filtering serves to get rid of non-interesting items \cite{Shneiderman1996}. To this end, we allow for the filtering of those nodes and/or edges whose frequencies are below given thresholds that can be configured using two scroll bars and/or two text boxes on the left side.

By increasing the value of the thresholds (only very frequent states and events will be shown), the automaton becomes more readable and therefore the user can more easily identify meaningful routes and important states from a pedagogical point of view.

The visualization, in addition to the global view of the automaton, has two views that filter the information in the model used to construct the graph per dates and per student. These views are accessed through the top menu. The per date view presents a model built only with data recorded within a range of dates specified by the user. This functionality can help, for example, to compare the behavior of students belonging to different academic years, or the behavior before and after a specific tutoring action.

The per student view presents the sequence of events performed by a specific student. This view allows the teacher to analyze the individual behavior of each student.

\subsubsection{Details-on-demand}
According to Schneiderman details of an object must be provided whenever the user requires it \cite{Shneiderman1996}. Following this recommendation we allow the user to obtain details of each state or event through mouse clicks. These details include identifiers, frequencies, zone the object belongs to, descriptions, tutoring messages, etc.

\subsubsection{Relate}
\begin{figure}[h]
 \centering
\includegraphics[width=0.45\textwidth,height=2cm]{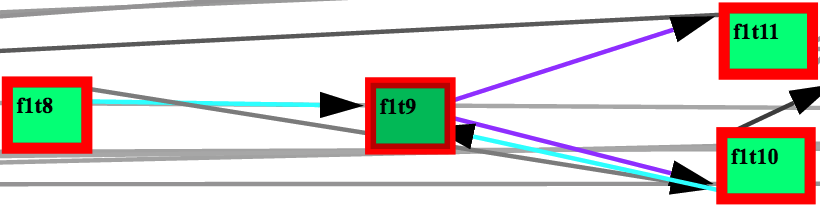}
 \caption{Example of relations between events and states.} 
 \label{fig:visor_coloresevt2}
\end{figure}

we allow the user to highlight relevant relationships between events and states. For this, when a node is selected, its edges are painted in colors (Figure \ref{fig:visor_coloresevt2})), differentiating the outcoming edges (in purple) from the incoming edges (in light blue).

The design decisions explained in this section were implemented in a web tool o viewer. The implementation details of this tool are documented in two undergraduate \cite{tfgAlvaro} and master \cite{tfmKeli} theses. This viewer can be executed in \url{http://138.100.14.99/SBPViewer}.

\section{Viewer applications} \label{sec:examples}
As mentioned above, this viewer is a tool that  instructors can use to adapt their teaching taking into account the behavior of students in the virtual environment. In addition, it can serve to identify those steps in the training process in which the tutoring strategy of an Intelligent Tutoring System could be improved.

Below we will describe three examples that illustrate how this viewer can assist an instructor in some teaching tasks. The 87 student logs that are visualized in these examples are real and have been taken from a virtual laboratory for teaching biotechnology \cite{Rico2012,Ricojucs2017,Ramirez2017}. This laboratory is an educational 3D virtual environment in which students carry out a practical assignment composed of around 120 actions, such as adding a chemical to a mix or turning on a machine. 

\subsection{Grading students}
This example aims to demonstrate the utility of the viewer to assist an instructor in the assessment of a student.

First we can use the viewer to find out how many students have finished the process without errors. To obtain this result we have to perform the following interactions in the viewer:
\begin{enumerate}
\item Open the global view of the model.
\item Write the name of the last action of the process (f3t61), select among the auto-completion results the state that belongs to the correct flow zone, and press the search button (zoom).
\item Click on the focused state (details on demand).
\end{enumerate}

Then, the viewer presents the detailed information of the state, in this case, a frequency of 1.47\% corresponding to 1 of 87 students.

\begin{figure}[b]
 \centering
\includegraphics[width=0.98\textwidth]{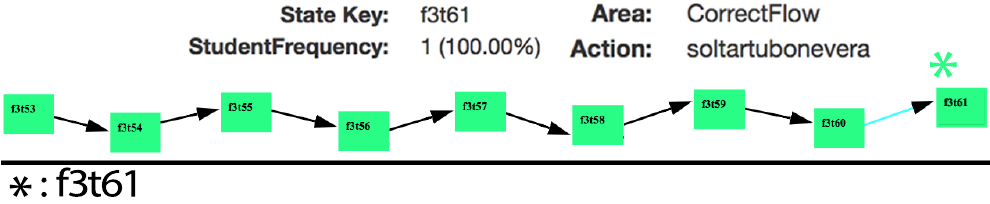}
 \caption{Correct states sequence of the student 22 in the biotechnology virtual lab.} 
 \label{fig:visor_student22}
\end{figure}

Let us suppose that the only student who successfully completed the exercise is the number 22. This is the only one in which the state f3t61 (state with green asterisk) was found in the correct flow zone (see Figure \ref{fig:visor_student22}). Note that otherwise, if this state is found in the relevant errors zone, this mean that the practical assignment was completed but with some errors.

For the other students the instructor may analyze how many and what kind of mistakes they made to give them a grade. This analysis is performed by navigating through the automaton and examining the details of the error states.

Let us focus on two student logs (for students 13 and 27), which committed some major mistakes and for this reason they would not deserve the same grade as student 22.

\begin{sidewaysfigure}
\centering
\textcolor{white}{\rule{0.75\textheight}{0.5\textheight}}
\includegraphics[width=\textwidth]{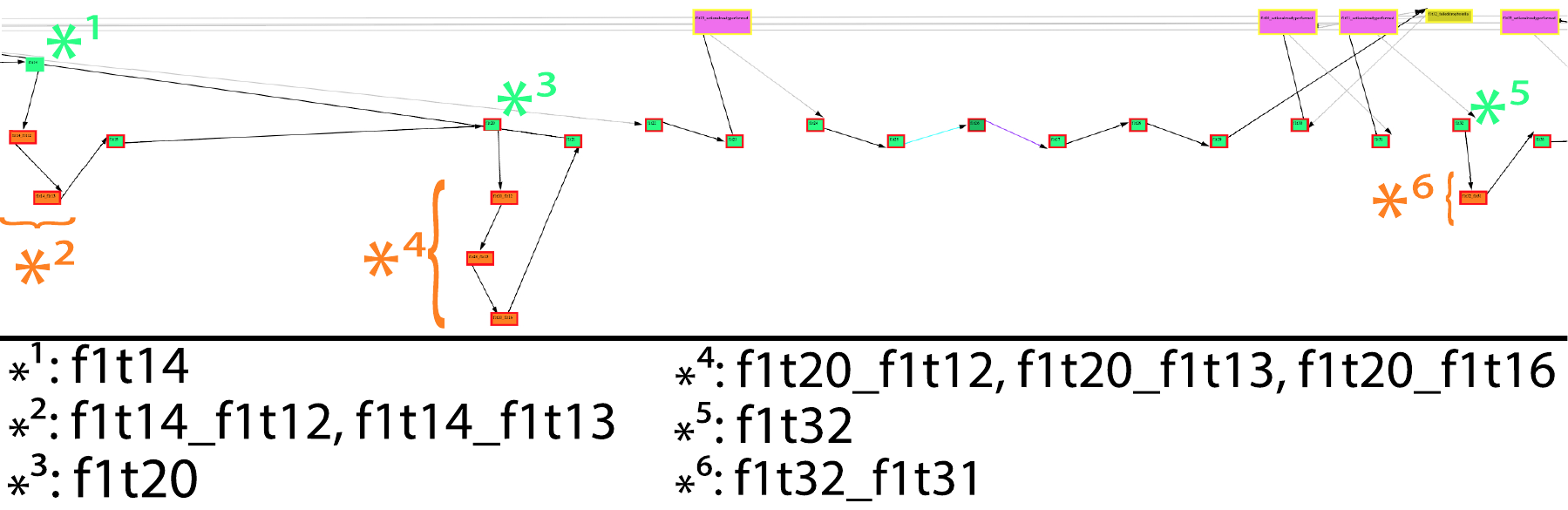}
 \caption{States sequence of student 13 in the biotechnology virtual lab.} 
 \label{fig:visor_student13}
\end{sidewaysfigure}

The states sequence of student 13 (Figure \ref{fig:visor_student13}) is correct until state f1t14 (green asterisk 1, action adjust pH), in which the automatic tutor detected that the student did not perform previously two actions (error states related to orange asterisk 2). Later, in action f1t20 (green asterisk 3), the automatic tutor found additional errors, also produced by a non executed action (error states related to orange asterisk 4). Later on, this student makes one more mistake (orange asterisk 6) produced by not turning on the laminar flow cabinet. This last error is not as  important as the previous errors that are the result of not adding substances to a mixture.

The instructor can easily visualize that this student only committed these errors in the whole process, and only in the first phase of the process. From this, it can be inferred that the student learned from his failures (which are explained to the student at the end of each phase) and improved in studying the actions necessary for the subsequent phases of the exercise.

\begin{figure}[h]
 \centering
 \subfloat[Errors in phase 1]{\label{fig:visor_student27_1}
\makebox[0.5\textwidth]{\includegraphics[width=0.2\textwidth,height=4cm]{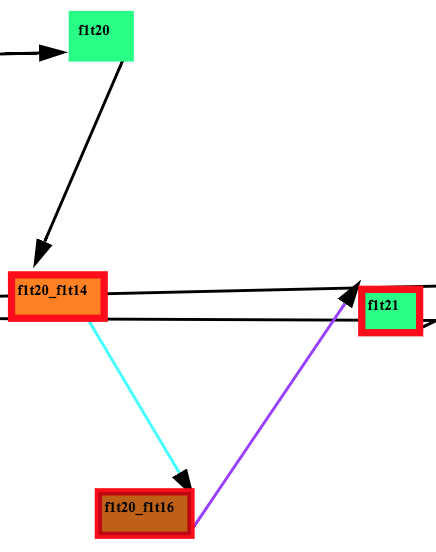}}}
\hfill
\subfloat[Errors in phase 2]{\label{fig:visor_student27_2}
\makebox[0.24\textwidth]{\includegraphics[width=0.1\textwidth,height=6cm]{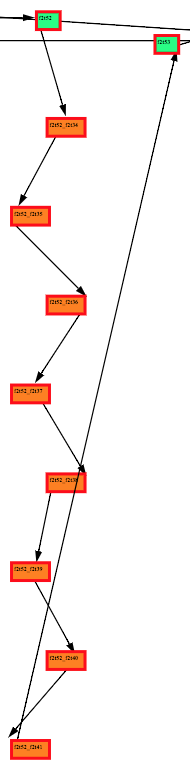}}}
\subfloat[Errors in phase 3]{\label{fig:visor_student27_3}
\makebox[0.24\textwidth]{\includegraphics[width=0.1\textwidth,height=5cm]{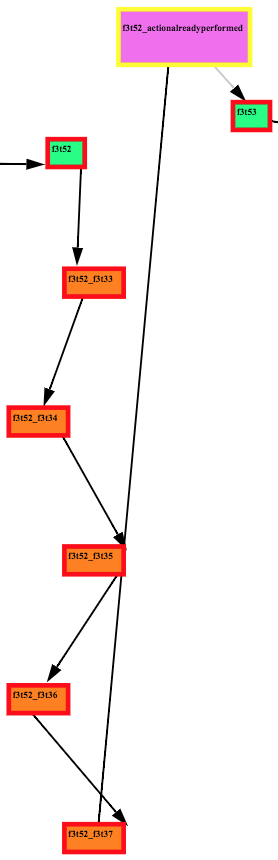}}}
\caption{States sequence of student 27 in the biotechnology virtual lab.} 
 \label{fig:visor_student27}
\end{figure}

Unlike student 13, the student 27 (see Figure \ref{fig:visor_student27}) did not learn from his/her previous mistakes and committed several errors in all phases of the learning process. Additionally, all these errors have a similar nature, because all of them are related to not adding certain components to different mixtures. In the global view of the model, the instructor can search the errors that have been committed by this student and analyze if those errors are frequent among all students. In this case, only two errors exceed 60\% in frequency, but the others are below 30\%.

\subsection{Comparing the behavior of students of different clusters}\label{sec:comparingclusters}
Another useful application of this viewer is to compare the action sequences performed by students of different clusters. In this way, a different tutoring strategy can be defined for the students of each cluster. For this we can open the global view of the model in several browser tabs (as many  as clusters), and visualize a different cluster in each one.

\begin{sidewaysfigure}
 \centering
\textcolor{white}{\rule{0.75\textheight}{0.5\textheight}}
 \subfloat[First part of the automaton of the cluster 0.]{\label{fig:visor_ejcompzoomclu0}
\includegraphics[width=\textwidth,height=10cm]{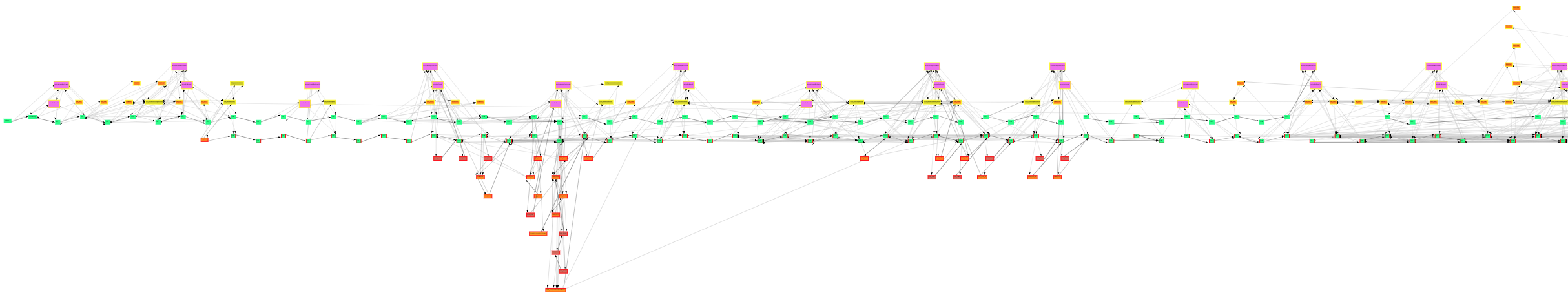}}
 \par
 \subfloat[First part of the automaton of the cluster 1.]{\label{fig:visor_ejcompzoomclu1}
\includegraphics[width=\textwidth,height=2cm]{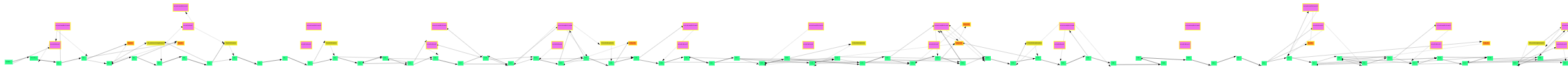}}
\caption{Global view of the automaton in each cluster.} 
 \label{fig:visor_ejcompzoom}
\end{sidewaysfigure}

One of the first steps in this comparison may focus on the global view of the states sequence in each cluster, offering a general overview of the students' behavior in each group (see Figure \ref{fig:visor_ejcompzoom}). In Figure \ref{fig:visor_ejcompzoom} we can see the differences between the first section of the two automata corresponding to two clusters of student logs. The automaton in cluster 0 has many states in the relevant error zone, whereas the automaton in cluster 1 does not. From this fact we can conclude that the students in cluster 1 had a quite better performance than the ones in cluster 0.

\begin{figure}[b]
 \centering
\includegraphics[width=0.85\textwidth,height=0.3\textheight]{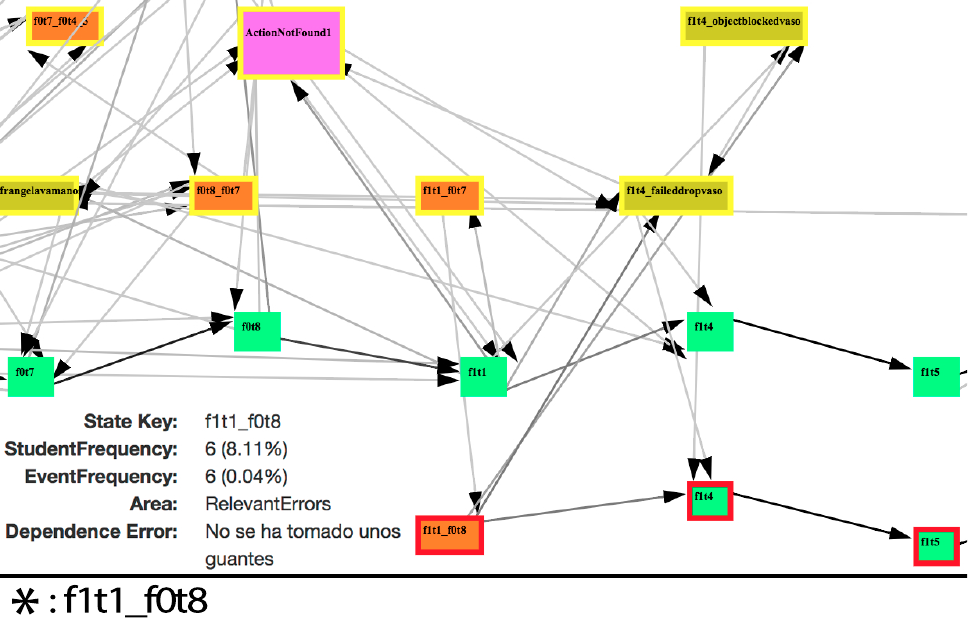}
 \caption{First error in the automaton of cluster 0.} 
 \label{fig:visor_ejcomf1t1clu0}
\end{figure}

\begin{figure}[h]
 \centering
\includegraphics[width=0.5\textwidth,height=0.5\textheight]{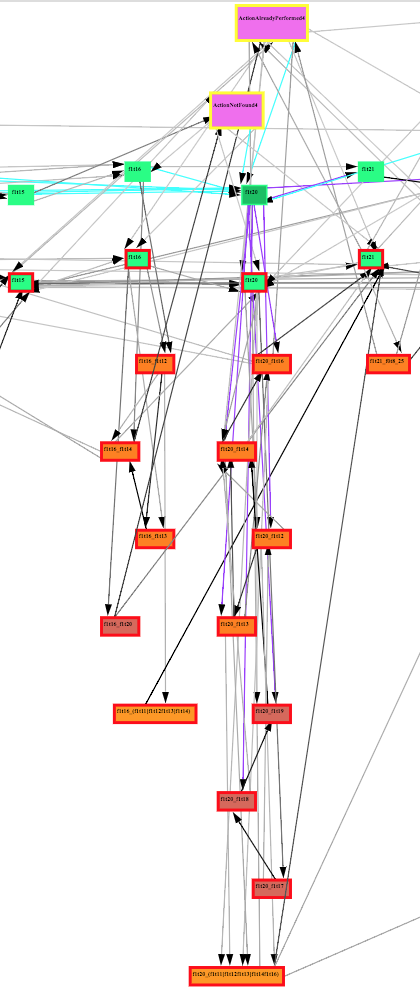}
 \caption{First error sequences in the automaton of cluster 0.} 
\label{fig:visor_ejcompf1t20clu0}
\end{figure}

Focusing on the first error in the first automaton, it can be seen that very few students (around 8\%) forgot to take rubber gloves before obtaining a beaker (error state f1t1\_f0t8 with orange asterisk in Figure \ref{fig:visor_ejcomf1t1clu0}). The frequency of this error state is low and this error does not exist in the automaton of the other group. For these reasons, this error is not relevant to draw significant conclusions.

In contrast, in the two leftmost fail sequences of the automaton in the same cluster (see Figure \ref{fig:visor_ejcompf1t20clu0}), most of the error states have frequencies greater than 60\%. However, since there are no such errors in the other group of students, it can be said that these errors are not necessarily associated with the method used to teach the subject. To mitigate this type of error in future students, the automatic tutoring strategy for that cluster should be improved in the states located just before these errors.

\begin{figure}[!htb]
 \centering
 \subfloat[Errors in cluster 0.]{\label{fig:visor_ejcompf2t52clu0}
\includegraphics[width=0.3\textwidth,height=0.45\textheight]{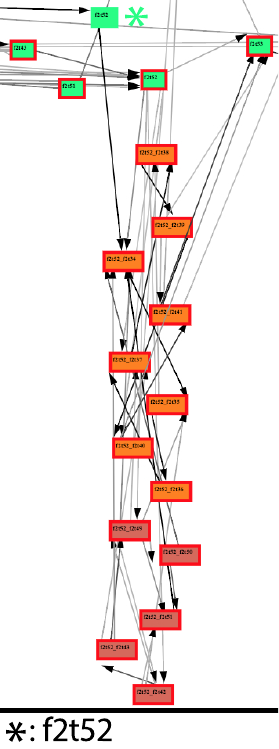}}
\quad
 \subfloat[Errors in cluster 1.]{\label{fig:visor_ejcompf2t52clu1}
\includegraphics[width=0.4\textwidth]{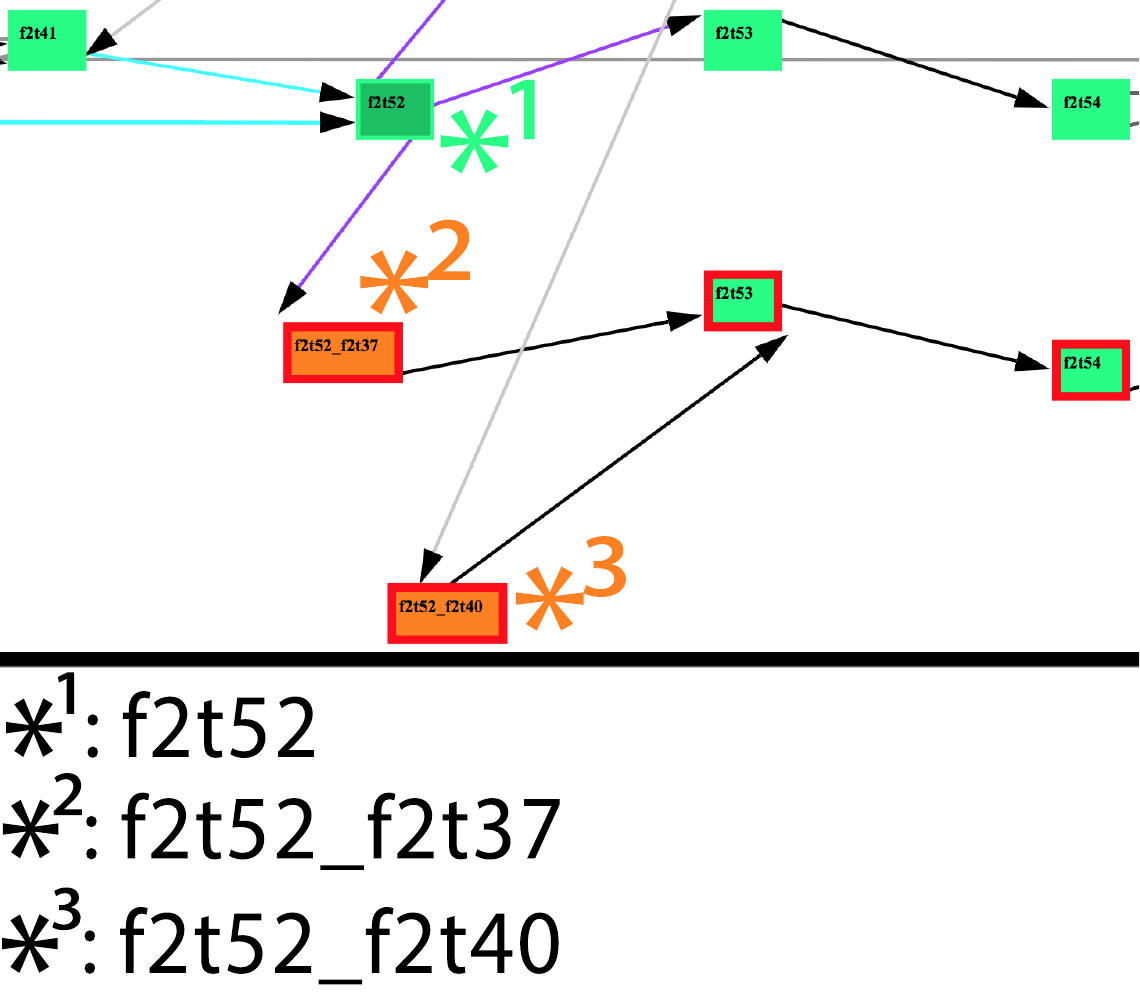}}
\caption{Comparison of errors at the end of the mix in phase 2 in the automatons of the two groups of students.} 
 \label{fig:visor_ejcompf2t52}
\end{figure}

The first relevant errors in cluster 1 (f2t52\_f2t37 and f2t52\_f2t40) appeared in the state f2t52 of phase 2 (see Figure \ref{fig:visor_ejcompf2t52}(b)) , where the automatic tutor is validating the components of a mixture. According to the automaton, half of the students forgot to add at least one of the required components. If we search for this state in the other cluster, we can see that, in addition to these same errors, there are others that were produced because some students forgot to add other components to the mixture and/or added some non-required components (Figure \ref{fig:visor_ejcompf2t52}(a)).

As there are some errors that appear with a high frequency in both clusters, it makes sense to conclude that the professor needs to improve his/her teaching methodology so that students can assimilate the concepts associated with these errors in a better way.

\begin{figure}[!htb]
 \centering
 \subfloat[Final states sequence in cluster 0.]{\label{fig:visor_ejcompf3t61clu0}
\includegraphics[width=0.45\textwidth]{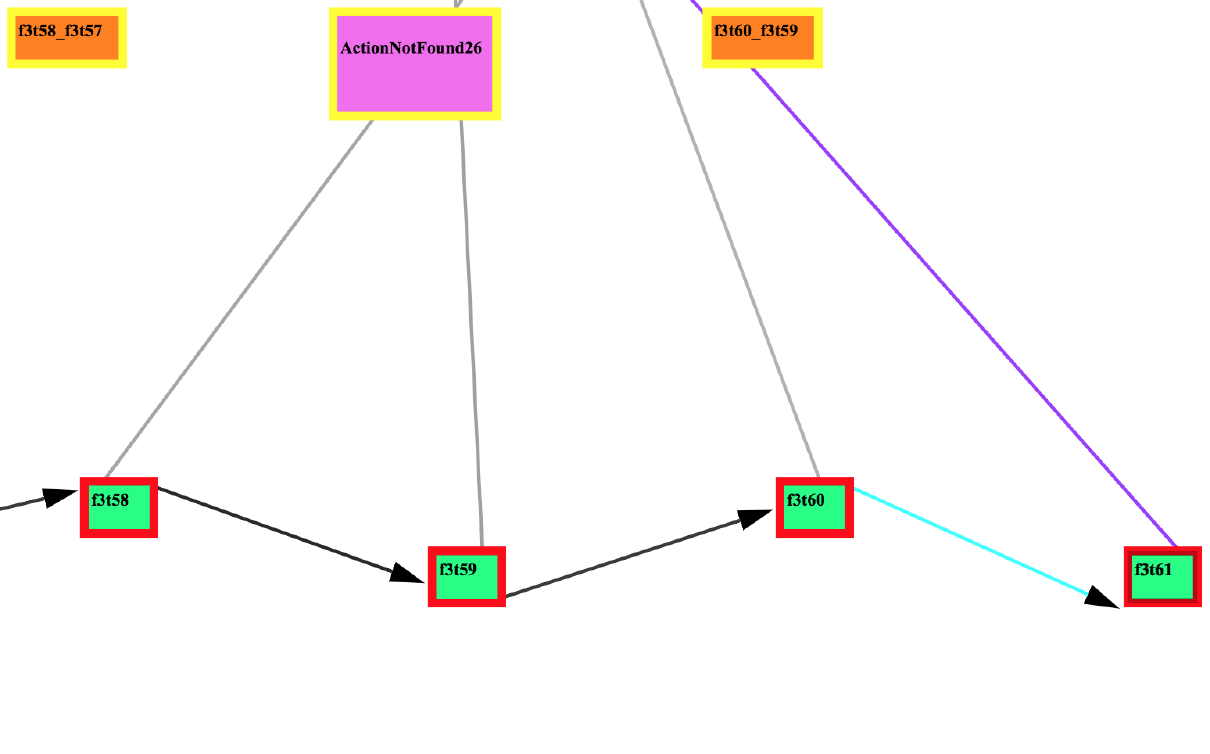}}
\quad
 \subfloat[Final states sequence in cluster 1.]{\label{fig:visor_ejcompf3t61clu1}
\includegraphics[width=0.45\textwidth]{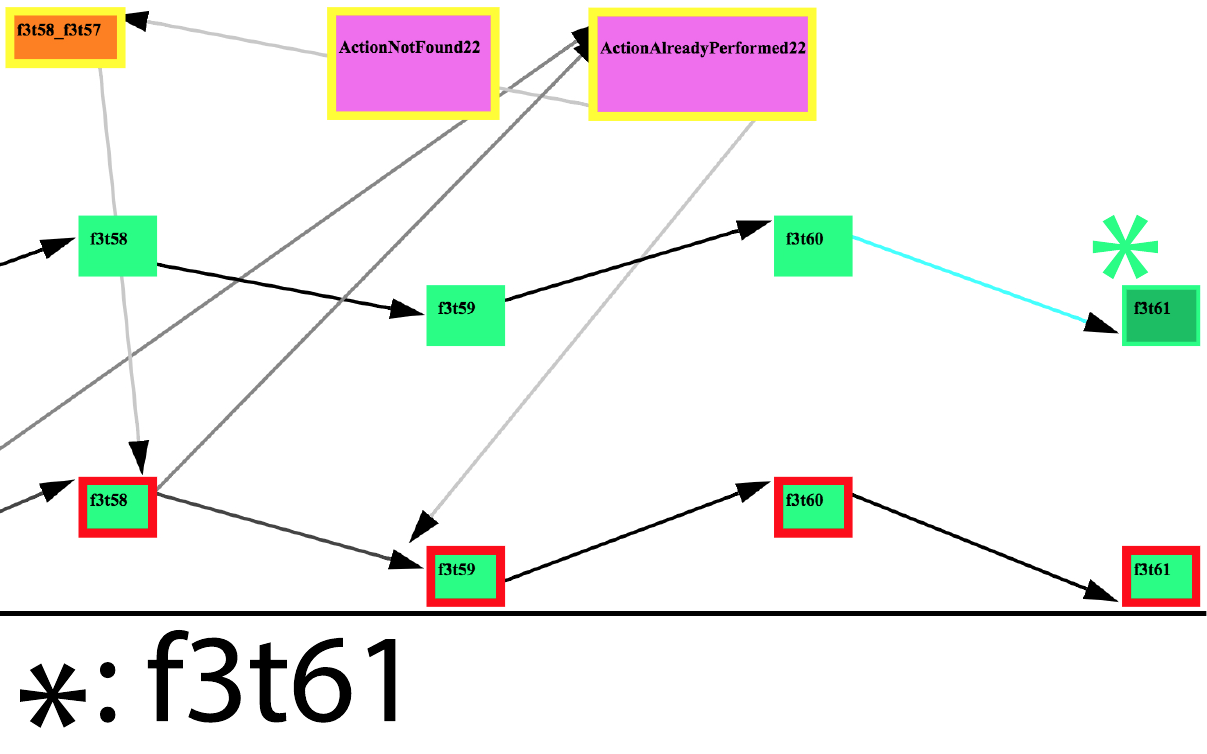}}
\caption{Final states sequence of the automatons of the two groups of students.} 
 \label{fig:visor_ejcompf3t61}
\end{figure}

Finally, if we search for the last state in the correct flow (state f3t61) in both automata, we can conclude (see Figure \ref{fig:visor_ejcompf3t61}) that none of the students in cluster 0 correctly completed the exercise, unlike cluster 1.

\subsection{Finding common mistakes}
One of the most useful application of the viewer is to discover common states that occur in the training process, such as errors.  Using the global view of the model, states (and events) with low frequencies can be filtered out to display only the most common states. Subsequently, with the panning function, we can navigate through the automaton to find the states associated with pedagogically interesting errors. Then, starting from the desired state, the viewer allows users to analyze the sequence of actions that students have taken to reach it. 

\begin{figure}[!htb]
 \centering
\includegraphics[width=0.45\textwidth,height=0.25\textheight]{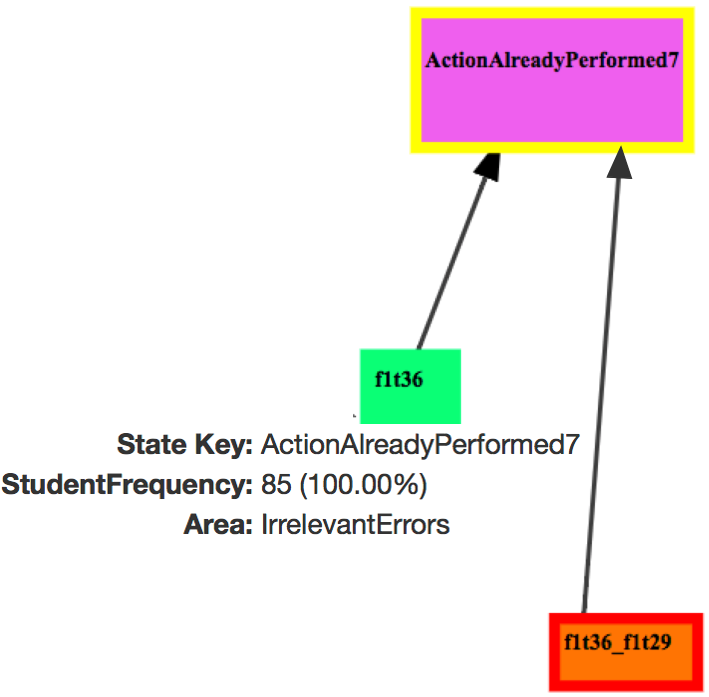}
 \caption{Frequent irrelevant error without interest for the instructor.}
\label{fig:visor_ejerrores_f1t37_alreadyperformed}
\end{figure}

Figure \ref{fig:visor_ejerrores_f1t37_alreadyperformed} presents an error in the irrelevant zone. This figure indicates that a high number of students committed the error of repeating an action (in this case sterilizing the material), either from a correct state or after having made another error.

The fact that many students try to redo the action of sterilizing the material does not probably mean that they did not assimilate the importance of the sterilization, but they had trouble with the interaction with the environment. For this reason, this error was not considered pedagogically interesting to the instructor.

\begin{figure}[!htb]
 \centering
\includegraphics[width=0.8\textwidth]{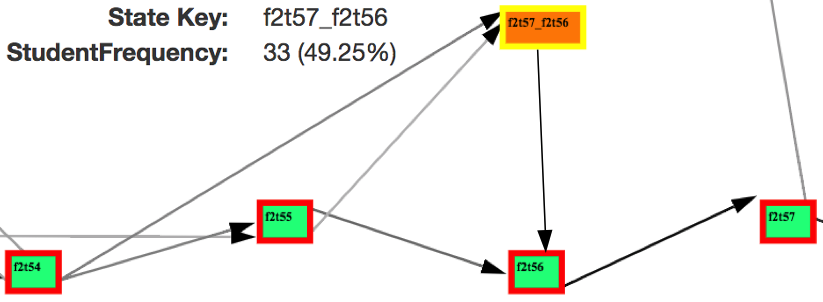}
 \caption{Frequent irrelevant error interesting for the instructor.} 
\label{fig:visor_ejerrores_f2t57_f2t56}
\end{figure}

Another common mistake discovered in student logs is depicted in Figure \ref{fig:visor_ejerrores_f2t57_f2t56}. This error arises when students attempt to initiate the PCR process without having previously introduced the tube with the mix into the machine.

Unlike the previous example, this error was indeed pedagogically interesting to the instructor, as it might reveal that many students are not assimilating correctly the functionality of a PCR machine.

\begin{figure}[!htb]
 \centering
\includegraphics[width=0.7\textwidth]{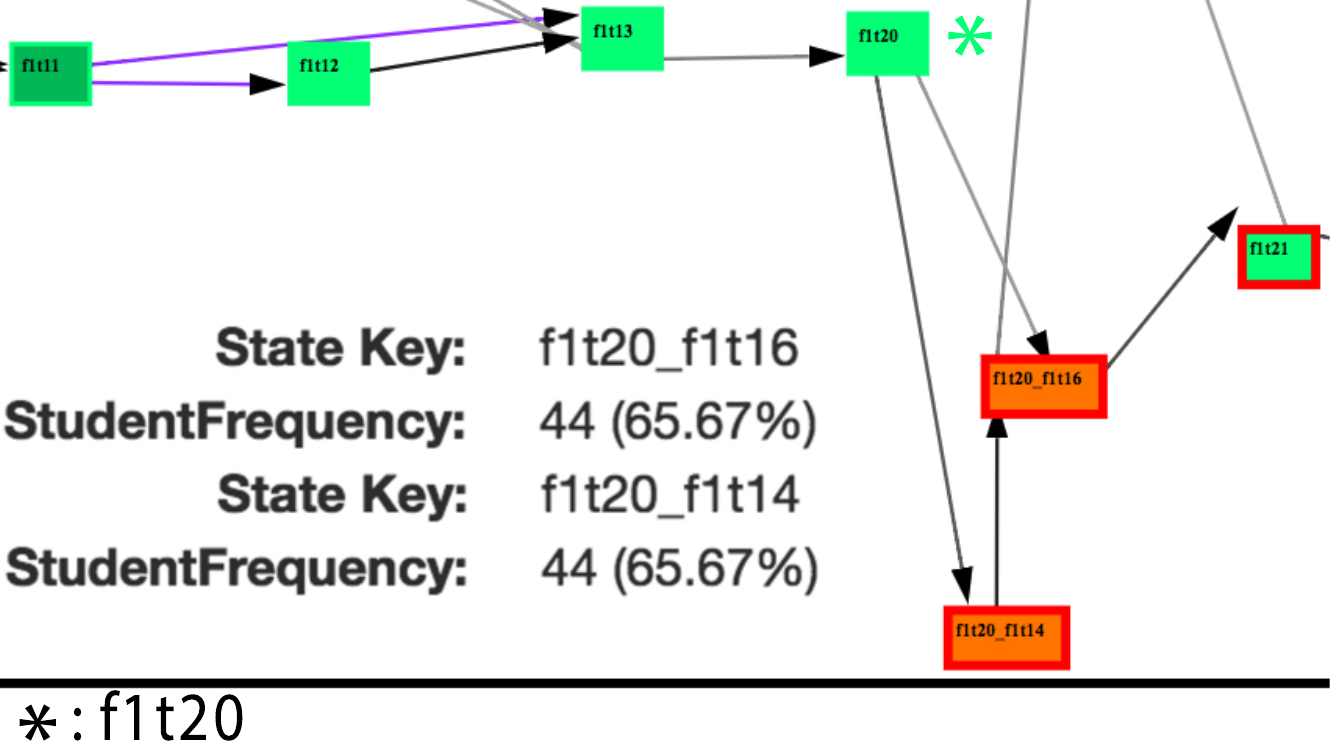}
 \caption{Frequent relevant errors.} 
\label{fig:visor_ejerroresrelevantes}
\end{figure}

In the last example, presented in Figure \ref{fig:visor_ejerroresrelevantes}, we can see two relevant errors with a frequency greater than 60\%. These errors occur when the automatic tutor validates the ``turn off the mixer'' action (f1t20). One is generated by not adding \textit{bacto agar} to the mix, whereas the other is the result of not adjusting the pH of the mix.

These kinds of mistakes are pedagogically interesting for the instructor, and thanks to the viewer, he/she can discover the steps of the process that students assimilated in a less effective way, and he/she can enhance the way of teaching the concepts related to these steps. Also, the automatic tutoring strategy could be modified to give students a hint in this step, for example a message asking them to check in their notes what the components of the mix are.

\section{Evaluation of the viewer utility} \label{sec:validation}
The objective of this evaluation was to verify the utility of the viewer to improve the effectiveness of the virtual environment.  First, the viewer was used to find a set of typical errors. Then three kinds of changes associated with these typical errors were performed in the virtual laboratory: changes in the tutoring strategy provided by an automatic tutor; changes in the chemical names; and minor changes in 3D objects. Subsequently, we compared the frequencies of these typical errors in the first version of the virtual laboratory with the frequencies of these same errors in the modified version of the virtual laboratory to evaluate the influence of the changes.

As we will explain below, the tutoring strategy was improved essentially by adding new hints or changing some existing tutoring messages. However, for most of the errors, we did not want to give very specific hints, because we wanted to encourage students to discover by themselves how to perform certain steps of the exercise. Hence, most of the new hints suggest students to review the PROTOCOL book that is available in the virtual world. These new hints, in turn, helps to foster a good practice at real laboratories, the use of the PROTOCOL book.

\subsection{Procedure}

In years 2013, 2014 and 2015, 85 students used the biotechnology virtual laboratory \cite{Rico2012,Ricojucs2017,Ramirez2017} to complete a practical assignment. The automatic tutor of this virtual lab collected and classified the events that all these students generated during the practical assignment, and the 85 student logs were used to build a collective student model. The resulting graph of this model (without clustering) contained 532 states and 2778 edges.

Then, an instructor of the subject that includes this practical assignment with the help of the developers of the viewer identified some critical zones of the process, i.e. those in which these students committed more mistakes, by using the viewer. The errors found are enumerated in Table \ref{tab:errores_pre2016}. The first column describes the nature of the error found; the second column specifies the action in which the automatic tutor generates the error events (when validating the preconditions of this action); the third column enumerates the list of error events associated with the general error; and the last column shows the identifier of the change that has been performed to reduce the frequency of the error. Theses changes are detailed in Tables \ref{tab:cambios_tutoria}, \ref{tab:cambios_quimicos} and \ref{tab:cambios_lab}.

\newcounter{rowno}
\setcounter{rowno}{0}
\begin{table}[h]
\centering
\caption{Common errors found thanks to the viewer.}
\label{tab:errores_pre2016}
\begin{tabularx}{0.98\textwidth}{X|l|l|c}
\hline
\textbf{Mistake} & \textbf{Action} & \textbf{Error} & \textbf{Change}\\
\hline
The pH of the mixture was not adjusted & f1t20 & f1t14 & \ref{item:cambio1}\\\hline
One or more of the components of the growing medium were not added and/or they were not added in the correct order & f1t20 & f1t12, f1t13, f1t16 &\ref{item:cambio2}\\\hline
The autoclave was not found & f1t25 & faileddropbotella & \ref{item:cambio3} and \ref{item:cambio13}\\\hline
One or more of the components of the PCR mix were not added and/or they were not added in the correct order & f2t52 & \begin{tabular}[t]{@{}l@{}}f2t34, f2t35, f2t36,\\ f2t37, f2t38, f2t39,\\ f2t40, f2t41\end{tabular} &\ref{item:cambio4}\\\hline
One or more of the components of the ligation mixture were not added and/or they were not added in the correct order & f3t52 & \begin{tabular}[t]{@{}l@{}}f3t33, f3t34, f3t35,\\ f3t36, f3t37\end{tabular} & \ref{item:cambio5}\\\hline
Agar was not added to the growing medium & f1t20 & f1t16 & \ref{item:cambio7} and \ref{item:cambio12}\\\hline
Polymerase Buffer was not added to PCR mix & f2t52 & f2t37 & \ref{item:cambio8}\\\hline
Taq Polymerase Enzyme was not added to the PCR mix. & f2t52 & f2t40 & \ref{item:cambio9}\\\hline
Ligase Buffer was not added to the ligation mix. & f3t52 & f3t35 & \ref{item:cambio10}\\\hline
PCR product was not added to the ligation mix. & f3t52 & f3t34 & \ref{item:cambio11}\\
\hline                   
\end{tabularx}
\end{table}

Once these problematic actions were identified, some tutoring messages were changed (see Table \ref{tab:cambios_tutoria}) and/or some additional help was added within the virtual environment (see Tables \ref{tab:cambios_quimicos} and \ref{tab:cambios_lab}).

\begin{table}[h]
\centering
\caption{Changes in tutoring messages.}
\label{tab:cambios_tutoria}
\begin{tabularx}{0.98\textwidth}{>{\refstepcounter{rowno}\therowno\label{item:cambio\therowno}}c|X|X}
\hline
\multicolumn{1}{c|}{\textbf{Nº}} &\textbf{Previous message} & \textbf{New message}\\
\hline
& There was no message reminding the student to adjust the pH & Remember to adjust the pH before adding the Agar\\\hline
& Touch the mixer to turn it on & Touch the mixer to turn it on. Review the composition of the medium in the PROTOCOL\\\hline
& There was no message to help the student find the autoclave & Look for the AUTOCLAVE in another room, this is like a pressure cooker with the size of a washing machine\\\hline
& There was no message reminding the student to check which components form the PCR mix & Review the composition of the PCR mix in the PROTOCOL \\\hline
& There was no message reminding the student to review which components form the ligation mix & Review the composition of the ligation mix in the PROTOCOL\\\hline
& Only the list of relevant errors was presented & It was also added: CONCLUSION: TO AVOID ERRORS REVIEW THE PROTOCOL\\
\hline                                
\end{tabularx}
\end{table}

\begin{table}[h]
\centering
\caption{Changes in chemical names.}
\label{tab:cambios_quimicos}
\begin{tabularx}{0.98\textwidth}{>{\refstepcounter{rowno}\therowno\label{item:cambio\therowno}}c|X|X}
\hline
\multicolumn{1}{c|}{\textbf{Nº}} &\textbf{Previous name} & \textbf{New name}\\
\hline
 & Bactoagar & Agar                            \\\hline
 & Polymerase Buffer & Polymerase Enzyme Buffer\\\hline
 & Taq Polymerase Enzyme & ADN Polymerase      \\\hline
 & Ligase Buffer & Ligase Enzyme Buffer     \\ \hline
 & PCR product & PCR amplified DNA  \\
\hline           
\end{tabularx}
\end{table}

\begin{table}[h]
\centering
\caption{Other changes in the virtual lab.}
\label{tab:cambios_lab}
\begin{tabularx}{0.98\textwidth}{>{\refstepcounter{rowno}\therowno\label{item:cambio\therowno}}c|X}
\hline
\multicolumn{1}{c|}{\textbf{Nº}} &\textbf{Change Description}\\
\hline	
&The chemical Agarose was removed, because its name is too similar to Agar and this confused students. \\\hline
& The Autoclave name was added on the 3D object that represents it\\
\hline
\end{tabularx}
\end{table}

Subsequently, 17 students who studied the same subject in 2016 used the virtual laboratory with the modifications.

For the analysis of collected data, therefore, we considered two periods:
\begin{description}
\item[Period 1]\ Students of years 2013, 2014 and 2015 (before changes).
\item[Period 2]\ \ Students of 2016 (after changes).
\end{description}

We verified statistically that these two groups were samples belonging to the same universe. For this, the final grades obtained in the subject of Biochemistry and Biotechnology (for which the virtual laboratory was used) were compared. The non-parametric U test from Mann-Whitney was used because data did not meet the normality test, providing a level of significance of 0.142. Therefore we concluded that the distribution of grades was the same for the two periods, so both samples belong to the same universe and can be compared.

\subsection{Results}
For the analysis, we calculated the differences between the frequencies of the error states produced by the students of the first period, and those committed by the students of the second period. In addition, we differentiated the error events that may have been affected by a change from those who may not.

We compared the distributions of the frequency differences of the errors associated with some change with the differences in the frequencies of the errors that are not associated with any change. For this, we posed the following hypotheses:

\begin{description}
\item[$H_0$] Variations in error frequencies do not depend on a change in the training environment.
\item[$H_1$] Variations in error frequencies actually depend on a change in the training environment.
\end{description}

By performing a t-test on the above-mentioned data, we obtained a level of significance less than 0.001 (0.0000031). Therefore we can refuse $H_0$, i.e., the changes that have occurred in the error frequencies actually depend on the improvements that have been made in the virtual laboratory.

Tables \ref{tab:frecuencias_errores1}, \ref{tab:frecuencias_errores2} and \ref{tab:frecuencias_errores3} present the student frequencies for each error that has been affected by the aforementioned changes and for each period. Errors with frequencies lower than 30\% in the two periods have been removed from the tables, since errors with a relatively low frequency are not of interest for automatic tutoring. In addition, these tables highlight high frequencies and the highest differences with the range of colors defined as:

\begin{table}[h]
\centering
\begin{tabular}{ll}
\cellcolor[RGB]{221,235,247} & Frequencies between 25\% and 50\%\\
\cellcolor[RGB]{155,194,230}& Frequencies over 50\%
\end{tabular}
\end{table}

The first column of these tables shows the identifiers of the changes that have been made in the training environment. The second and third columns present the codes of the actions that generate the error events and the resulting error events, respectively.

Table \ref{tab:frecuencias_errores1} presents frequencies associated to changes in tutoring messages that give the student a general feedback.

\begin{table}[h]
\centering
\caption{Frequencies in error states associated with changes in general tutoring messages}
\label{tab:frecuencias_errores1}
\begin{tabular}{c|cc|cc|c}
\hline
\textbf{Change} &  \textbf{Action} & \textbf{Error} & \textbf{\begin{tabular}[c]{@{}c@{}}Freq.\\ 2013-2015\\ (\%)\end{tabular}} & \textbf{\begin{tabular}[c]{@{}c@{}}Freq.\\ 2016\\ (\%)\end{tabular}} & \textbf{Difference} \\
\hline
\multirow{4}{*}{\ref{item:cambio2}} & \multirow{3}{*}{f1t20} & f1t12 & 0.162 & \cellcolor[RGB]{221,235,247}0.353 & -0.191\\
 & & f1t13 & 0.147 & \cellcolor[RGB]{221,235,247}0.353 & -0.206\\
 & & f1t16 & \cellcolor[RGB]{155,194,230}0.662 & \cellcolor[RGB]{221,235,247}0.412 & \cellcolor[RGB]{221,235,247}0.250\\\cline{2-6} 
&\multicolumn{2}{r|}{\textbf{Average}} & 0.324 & 0.373 & -0.049\\
\hline
\multirow{9}{*}{\ref{item:cambio4}} & \multirow{8}{*}{f2t52} & f2t34 & \cellcolor[RGB]{221,235,247}0.456 & \cellcolor[RGB]{221,235,247}0.353 & 0.103\\
 & & f2t35 & \cellcolor[RGB]{155,194,230}0.529 & \cellcolor[RGB]{221,235,247}0.412 & 0.118\\
 & & f2t36 & \cellcolor[RGB]{221,235,247}0.485 & \cellcolor[RGB]{221,235,247}0.294 & 0.191\\
 & & f2t37 & \cellcolor[RGB]{155,194,230}0.588 & \cellcolor[RGB]{221,235,247}0.412 & 0.176\\
 & & f2t38 & \cellcolor[RGB]{155,194,230}0.574 & \cellcolor[RGB]{221,235,247}0.353 & 0.221\\
 & & f2t39 & \cellcolor[RGB]{155,194,230}0.574 & \cellcolor[RGB]{221,235,247}0.412 & 0.162\\
 & & f2t40 & \cellcolor[RGB]{155,194,230}0.647 & \cellcolor[RGB]{221,235,247}0.353 & \cellcolor[RGB]{221,235,247}0.294\\
 & & f2t41 & \cellcolor[RGB]{155,194,230}0.515 & \cellcolor[RGB]{221,235,247}0.471 & 0.044\\\cline{2-6} 
&\multicolumn{2}{r|}{\textbf{Average}} & 0.546 & 0.382 & 0.164\\
\hline
\multirow{6}{*}{\ref{item:cambio5}} & \multirow{5}{*}{f3t52} & f3t33 & \cellcolor[RGB]{221,235,247}0.397 & 0.176 & 0.221\\
 & & f3t34 & \cellcolor[RGB]{155,194,230}0.588 & 0.118 & \cellcolor[RGB]{221,235,247}0.471\\
 & & f3t35 & \cellcolor[RGB]{221,235,247}0.309 & 0.118 & 0.191\\
 & & f3t36 & \cellcolor[RGB]{221,235,247}0.309 & 0.118 & 0.191\\
 & & f3t37 & \cellcolor[RGB]{221,235,247}0.397 & 0.176 & 0.221\\\cline{2-6} 
&\multicolumn{2}{r|}{\textbf{Average}} & 0.400 & 0.141 & 0.259\\
\hline
\multicolumn{3}{r|}{\textbf{Average Total}} & 0.459 & 0.305 & 0.153\\
\hline
\end{tabular}
\end{table}

Table \ref{tab:frecuencias_errores2} shows frequencies associated to changes in specific tutoring messages. The change identifier \ref{item:cambio13} is included in this table, although it is not an improvement in a tutoring message, because it influences the same errors as change \ref{item:cambio3}.

\begin{table}[h]
\centering
\caption{Frequencies in error states associated with changes in specific tutoring messages}
\label{tab:frecuencias_errores2}
\begin{tabular}{c|cc|cc|c}
\hline
\textbf{Change} &  \textbf{Action} & \textbf{Error} & \textbf{\begin{tabular}[c]{@{}c@{}}Freq.\\ 2013-2015\\ (\%)\end{tabular}} & \textbf{\begin{tabular}[c]{@{}c@{}}Freq.\\ 2016\\ (\%)\end{tabular}} & \textbf{Difference} \\
\hline
\ref{item:cambio1} & f1t20 & f1t14 & \cellcolor[RGB]{155,194,230}0.662 & \cellcolor[RGB]{221,235,247}0.353 & \cellcolor[RGB]{221,235,247}0.309\\
\ref{item:cambio3} and \ref{item:cambio13} & f1t25 & faileddrop & \cellcolor[RGB]{221,235,247}0.382 & 0.118 & \cellcolor[RGB]{221,235,247}0.265\\
\hline
\multicolumn{3}{r|}{\textbf{Average}} & 0.522 & 0.235 & 0.287\\
\hline
\end{tabular}
\end{table}

Table \ref{tab:frecuencias_errores3} outlines frequencies of errors associated to changes that have been made in the terminology used in the virtual laboratory.

\begin{table}[h]
\centering
\caption{Frequencies in error states associated with changes in the terminology}
\label{tab:frecuencias_errores3}
\begin{tabular}{c|cc|cc|c}
\hline
\textbf{Change} &  \textbf{Action} & \textbf{Error} & \textbf{\begin{tabular}[c]{@{}c@{}}Freq.\\ 2013-2015\\ (\%)\end{tabular}} & \textbf{\begin{tabular}[c]{@{}c@{}}Freq.\\ 2016\\ (\%)\end{tabular}} & \textbf{Difference} \\
\hline
\ref{item:cambio7} and \ref{item:cambio12} & f1t20 & f1t16 & \cellcolor[RGB]{155,194,230}0.662 & \cellcolor[RGB]{221,235,247}0.412 & \cellcolor[RGB]{221,235,247}0.250\\
\ref{item:cambio8} & f2t52 & f2t37 & \cellcolor[RGB]{155,194,230}0.588 & \cellcolor[RGB]{221,235,247}0.412 & 0.176\\
\ref{item:cambio9} & f2t52 & f2t40 & \cellcolor[RGB]{155,194,230}0.647 & \cellcolor[RGB]{221,235,247}0.353 & \cellcolor[RGB]{221,235,247}0.294\\
\ref{item:cambio10} & f3t52 & f3t35 & \cellcolor[RGB]{221,235,247}0.309 & 0.118 & 0.191\\
\ref{item:cambio11} & f3t52 & f3t34 & \cellcolor[RGB]{155,194,230}0.588 & 0.000 & \cellcolor[RGB]{155,194,230}0.588\\
\hline
\multicolumn{3}{r|}{\textbf{Average}} & 0.559 & 0.259 & 0.300\\
\hline
\end{tabular}
\end{table}

Finally, although it is a change in a general tutoring message, change number \ref{item:cambio6} has been included in Table \ref{tab:frecuencias_errores4} instead of in Table \ref{tab:frecuencias_errores1}. This is because it does not prevent a unique error, but many relevant errors that may occur in the following phases. This improvement involves extending the message that is presented to the student at the end of each phase with an invitation to review the PROTOCOL book in the next steps of the exercise.

\begin{table}[h]
\centering
\caption{Frequencies in error states associated with message change at the end of each phase}
\label{tab:frecuencias_errores4}
\begin{tabular}{c|c|cc|c}
\hline
\textbf{Change} &  \textbf{Error} & \textbf{\begin{tabular}[c]{@{}c@{}}Freq.\\ 2013-2015\\ (\%)\end{tabular}} & \textbf{\begin{tabular}[c]{@{}c@{}}Freq.\\ 2016\\ (\%)\end{tabular}} & \textbf{Difference} \\
\hline
\multirow{3}{*}{\ref{item:cambio6}} & f1-f2-f3 & 0.368 & 0.059 & 0.309\\
 & f1-f2 & 0.147 & 0.412 & -0.265\\
 & f1 & 0.088 & 0.353 & -0.265\\
\hline
\end{tabular}
\end{table}

In this last table the row with the error \textit{f1-f2-f3} indicates the frequency of students who have made relevant errors in all three phases. Second row with error \textit{f1-f2} shows the frequency of those who have failed in the first two phases but not in the third. And the last row (\textit{f1}) outlines the frequency of those who were wrong only in the first phase.

\subsection{Discussion}
It is worth remarking that there is a general reduction in the frequency of errors in the second period with respect to the first period. This reduction is confirmed by the averages presented in the last row of the tables. In these averages it can be seen that the students in 2016 committed 15.3\% less errors with the new general tutoring messages; 28.7\% less errors with the new specific tutoring messages; and 30\% less errors thanks to the changes in terminology. As expected, these last two types of improvements have achieved a greater reduction in the frequency of errors, since they help students not to make mistakes in a more explicit way.

Despite these positive improvements in error reduction, the tutoring change number \ref{item:cambio2} has not helped students to make less mistakes in the creation of the growing medium, because the average of the difference for this change is -4.9\%. A possible solution to reduce errors in this part of the laboratory process could be to add more specific tutoring messages for actions such as ``add agar'' (f1t16), in which there is still a high frequency of mistakes in 2016. However, it should be the instructor of the subject who decides whether it is worthwhile to increase the level of detail of the hint in the tutoring strategy in those most problematic steps of the practical assignment, or, if otherwise, it is preferable to maintain this level of feedback and to spend more time to the explanation of the concept in his/her master class.

Table \ref{tab:frecuencias_errores4} reveals that the percentage of students that made mistakes in all the phases decreased in 2016. Therefore we can conclude that more students in 2016 than in the first period have taken the precaution of reading the PROTOCOL book more carefully, whenever they make mistakes in a previous phase.

In summary, improvements in the tutoring strategy and in the biotechnology lab have greatly reduced the number of errors committed by students in several critical actions (that were identified thanks to the viewer). Additionally, according to the statistical analysis detailed above, it was confirmed that this reduction in the number of errors in 2016 depends significantly on those improvements.

\section{Conclusions}\label{sec:conclusion}
This paper presents a viewer for the collective student model defined in \cite{Riofrio-Luzcando2017}, which draws an automaton on the screen and allows the user to interact with it. The visualization of the model has been designed following the recommendations to draw graphs proposed by Mazza \cite{Mazza2009} and the TTT taxonomy described by Shneiderman \cite{Shneiderman1996}.

This paper represents a noteworthy contribution to the stare of the art, because, to the best of our knowledge, there are not previous works that have addressed the visualization of data collected by tracking the actions performed by all the students in a virtual learning environment for procedural training in a way that help instructors to identify behavior patterns.

It has been illustrated how the viewer can help the instructors in: assigning qualifications to the students who have carried out the training process; comparing the behavior of students in different clusters; identifying common mistakes that point out knowledge gaps in the students; etc.

In order to check the utility of the presented viewer, we conducted an experiment. In this experiment, we showed how the viewer can serve to improve the automatic tutoring strategy and/or the teaching of a subject. For this, the instructor with the help of the developers of the viewer used it to find the most frequent errors that students committed performing an exercise in a 3D virtual laboratory in years 2013, 2014 and 2015. Once these errors were identified, we made changes in some tutoring messages and some elements of the virtual laboratory that could help prevent these errors. Then the exercise was repeated in 2016 revealing that the changes had a positive influence in the performance of the students.

The collective student model can be built from the student logs coming from any virtual environment for procedural training, as long as these logs are in the required format. Hence, the viewer may be used to visualize the student logs coming from any of these virtual environments. 

\section{Future Work}\label{sec:futurework}

One of the main limitations of the current viewer is its usability. Concerning this, we plan to introduce some changes in the user interface that help to improve the user experience.

A useful feature that could be added to the viewer would be the identification of subgraphs associated with pedagogically interesting patterns in the automaton. These patterns would be discovered by means of graph mining algorithms.

Additionally, an extended version of the viewer could help with the edition of the tutoring strategy associated with events or states in the automaton. For example, the viewer may support the modification of tutoring messages to prevent certain student's errors. 

This extended version could also include a new view so that users can compare the behavior of students of different clusters, like is explained in section \ref{sec:comparingclusters}. This view should present the automaton of each cluster in the same window to prevent the user from opening several tabs or windows of the browser to do this task.

So far the viewer has only been used by the instructor that was involved in the experiment. Besides, this instructor was assisted by the developers of the viewer while she was working with the viewer. Therefore, we would like to carry out tests with other instructors to evaluate the effectiveness with which they can perform certain tasks. This user evaluation might also serve to point out some issues that should be addressed to improve the usability of the viewer.

\section{Acknowledgments}
Riofr\'io would like to acknowledge financial support from the Ecuadorian Secretariat of Higher Education, Science, Technology and Innovation (SENESCYT).


\end{document}